\documentclass[10pt,twocolumn,letterpaper]{article}

\usepackage{cvpr}              

\usepackage{amsmath,amsfonts,bm}

\def\eqref#1{equation~\ref{#1}}

\def\1{\bm{1}}

\DeclareMathAlphabet{\mathsfit}{\encodingdefault}{\sfdefault}{m}{sl}
\SetMathAlphabet{\mathsfit}{bold}{\encodingdefault}{\sfdefault}{bx}{n}

\DeclareMathOperator*{\argmin}{arg\,min}

\usepackage{graphicx}
\usepackage{amsmath}
\usepackage{amssymb}
\usepackage{booktabs}
\usepackage{colortbl}
\usepackage{xcolor}
\usepackage{url}
\usepackage{mathtools}
\usepackage{microtype}
\usepackage{float}
\usepackage{bm}
\usepackage{amsfonts}
\usepackage{amsthm}
\usepackage{multirow}
\usepackage{adjustbox}
\usepackage{caption}
\usepackage{subcaption}

\usepackage{algorithm}
\usepackage{algorithmicx}
\usepackage{algpseudocode}
\usepackage{multirow}
\usepackage{adjustbox}
\usepackage{threeparttable}
\usepackage{pifont}
\usepackage{array}
\usepackage{enumitem}
\usepackage{bbding}
\usepackage{fontawesome}

\usepackage[pagebackref,breaklinks,colorlinks]{hyperref}

\usepackage[capitalize]{cleveref}
\crefname{section}{Sec.}{Secs.}
\Crefname{section}{Section}{Sections}
\Crefname{table}{Table}{Tables}
\crefname{table}{Tab.}{Tabs.}

\def \xx {{\bm{x}}}
\def \yy {{\bm{y}}}
\def \ee {{\bm{e}}}

\def \X  {\mathcal{X}}
\def \Y  {\mathcal{Y}}
\def \EE  {{\bm{E}}}
\def \citep {\cite}

\newcommand{\PreserveBackslash}[1]{\let\temp=\\#1\let\\=\temp}
\newcolumntype{C}[1]{>{\PreserveBackslash\centering}p{#1}}
\newcolumntype{R}[1]{>{\PreserveBackslash\raggedleft}p{#1}}
\newcolumntype{L}[1]{>{\PreserveBackslash\raggedright}p{#1}}

\def\cvprPaperID{1294} 
\def\confName{CVPR}
\def\confYear{2023}

\begin{document}


\title{Unlearnable Clusters: Towards Label-agnostic Unlearnable Examples}

\author{Jiaming Zhang\textsuperscript{\rm 1} \quad Xingjun Ma\textsuperscript{\rm 2}\thanks{Corresponding authors} \quad Qi Yi\textsuperscript{\rm 1} \quad Jitao Sang\textsuperscript{\rm 1,4}\footnotemark[1] \quad Yu-Gang Jiang\textsuperscript{\rm 2}\\
Yaowei Wang\textsuperscript{\rm 4} \quad \ Changsheng Xu\textsuperscript{\rm 3,4}\\
\textsuperscript{\rm 1}Beijing Jiaotong University \ \textsuperscript{\rm 2}Fudan University \ \textsuperscript{\rm 3}Chinese Academy of Sciences \ \textsuperscript{\rm 4}Peng Cheng Lab\\
}

\maketitle

\begin{abstract}
There is a growing interest in developing unlearnable examples (UEs) against visual privacy leaks on the Internet. UEs are training samples added with invisible but unlearnable noise, which have been found can prevent unauthorized training of machine learning models. 
UEs typically are generated via a bilevel optimization framework with a surrogate model to remove (minimize) errors from the original samples, and then applied to protect the data against unknown target models. 
However, existing UE generation methods all rely on an ideal assumption called \textbf{label-consistency}, where the hackers and protectors are assumed to hold the same label for a given sample. In this work, we propose and promote a more practical \textbf{label-agnostic} setting, where the hackers may exploit the protected data quite differently from the protectors. E.g., a $m$-class unlearnable dataset held by the protector may be exploited by the hacker as a $n$-class dataset.
Existing UE generation methods are rendered ineffective in this challenging setting.
To tackle this challenge, we present a novel technique called \textbf{Unlearnable Clusters} (UCs) to generate label-agnostic unlearnable examples with cluster-wise perturbations.
Furthermore, we propose to leverage Vision-and-Language Pre-trained Models (VLPMs) like CLIP as the surrogate model to improve the transferability of the crafted UCs to diverse domains.
We empirically verify the effectiveness of our proposed approach under a variety of settings with different datasets, target models, and even commercial platforms Microsoft {\tt Azure} and Baidu {\tt PaddlePaddle}. Code is available at \url{https://github.com/jiamingzhang94/Unlearnable-Clusters}.
\end{abstract}


\section{Introduction}\label{introduction}


\begin{figure}[t]
  \centering
  \includegraphics[width=\linewidth]{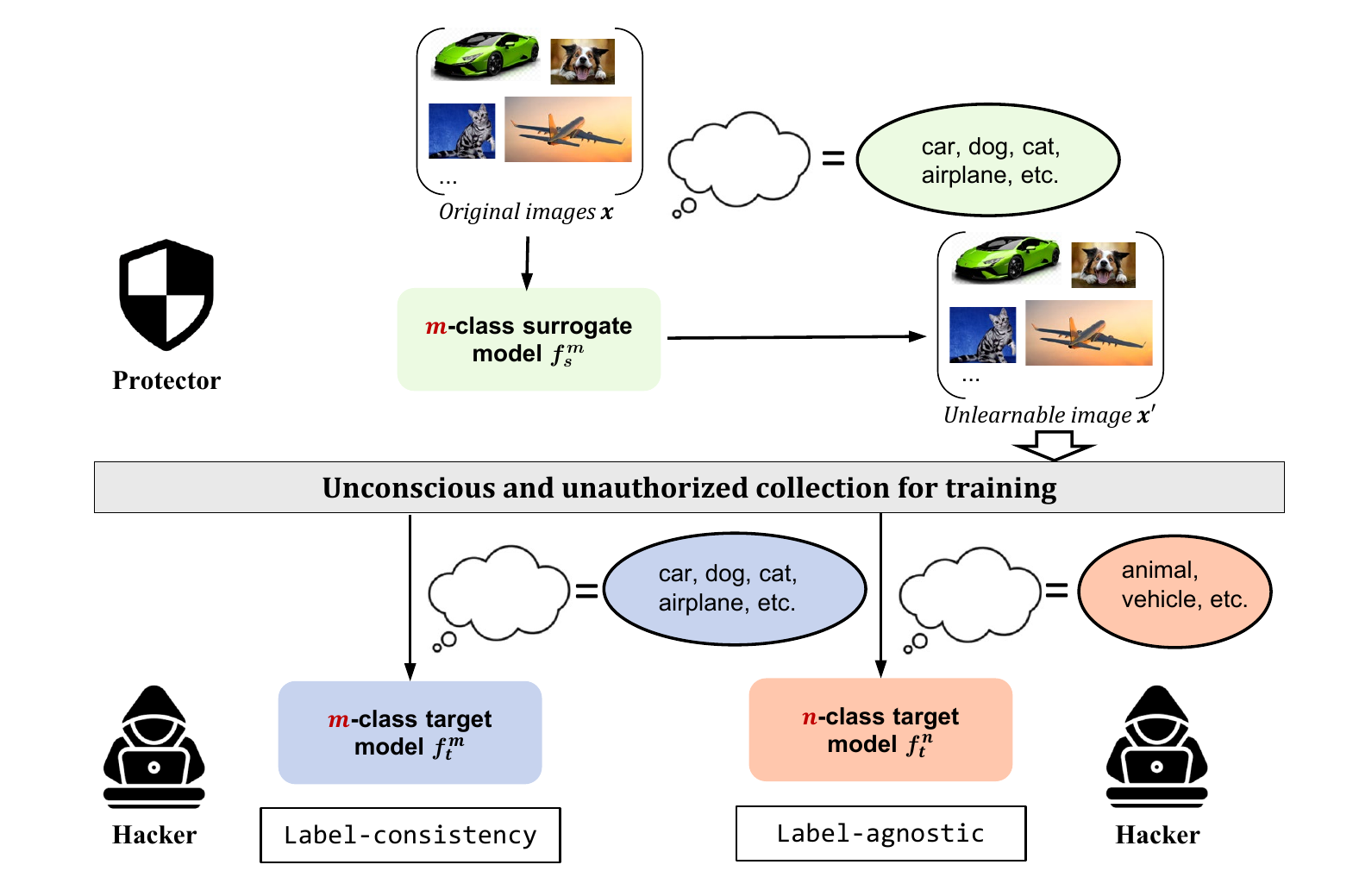}
  \caption{An illustration of two different data protection assumptions: \emph{label-consistency} vs. \emph{label-agnostic}, where the hacker exploits the protected data in different manners.}
  \label{fig_1}
\vspace{-10pt}
\end{figure}
While the huge amount of ``free" data available on the Internet has been key to the success of deep learning and computer vision, this has also raised public concerns on the unauthorized exploitation of personal data uploaded to the Internet to train commercial or even malicious models~\citep{hill2020secretive}. For example, a company named Clearview AI has been found to have scraped billions of personal images from Facebook, YouTube, Venmo and millions of other websites to construct a commercial facial recognition application~\citep{zhang2020adversarial}. 
This has motivated the proposal of \textbf{Unlearnable Examples} (UEs)~\citep{huang2020unlearnable} to make data \textbf{unlearnable} (or unusable) to machine learning models/services. Similar techniques are also known as availability attacks~\citep{biggio2018wild, yu2022availability} or indiscriminate poisoning attacks~\citep{he2022indiscriminate} in the literature. These techniques allow users to actively adding protective noise into their private data to avoid unauthorized exploitation, rather than putting our trust into the hands of large corporations.


The original UE generation method generates error-minimizing noise via a bilevel min-min optimization framework with a surrogate model \citep{huang2020unlearnable}. The noise can then be added to samples in a training set in either a sample-wise or class-wise manner to make the entire dataset unlearnable to different DNNs. It has been found that this method cannot survive adversarial training, which has been addressed by a recent method \citep{fu2022robust}. 
In this work, we identify one common assumption made by existing UE methods: \textbf{label-consistency}, where the hackers will exploit the protected dataset in the same way as the protector including the labels. This means that, for the same image, the hacker and protector hold the same label.
We argue that this assumption is too ideal, and it is possible that the hackers will collect the protected (unlearnable) samples into a dataset for a different task and label the dataset into different number of classes. 
As illustrated in Figure~\ref{fig_1}, an image can be labelled with different annotated labels (cat or animal), showing that a $m$-class (e.g., 10-class) unlearnable dataset may be exploited by the hacker as a $n$-class (e.g., 5-class or 20-class) dataset depending on its actual needs.
We term this more generic assumption as \textbf{label-agnostic} and propose a novel method \textbf{Unlearnable Clusters} (UCs) to generate more effective and transferable unlearnable examples under this harsh setting.

In Figure~\ref{fig_2} (a), we show that this more generic label-agnostic setting poses a unique transferability challenge for the noise generated by existing methods like Error-Minimizing Noise (EMinN)~\citep{huang2020unlearnable}, Adversarial Poisoning (AdvPoison)~\citep{fowl2021adversarial}, Synthetic Perturbations (SynPer)~\citep{yu2022availability} and DeepConfuse~\citep{feng2019learning}. This indicates that the protective noise generated by these methods are label-dependent and are rendered ineffective when presented with different number of classes. As such, we need more fundamental approaches to make a dataset unlearnable regardless of the annotations. To this end, we start by analyzing the working mechanism of UEs generated by EMinN, AdvPoison as they are very representative under the label-consistency setting. Through a set of visual analyses, we find that the main reason why they could break supervised learners is that the generated noise tends to disrupts the distributional uniformity and discrepancy in the deep representation space.
Uniformity refers to the property that the manifold of UEs in the deep representation space does not deviate much from that of the clean examples, while discrepancy refers to the property that examples belonging to the same class are richly diverse in the representation space.
Inspired by the above observation, we propose a novel approach called \textbf{Unlearnable Clusters} (UCs) to generate label-agnostic UEs using cluster-wise (rather than class-wise) perturbations.
This allows us to achieve a simultaneous disruption of the uniformity and discrepancy without knowing the label information.

Arguably, the choose of a proper surrogate model also plays an important role in generating effective UEs. Previous methods generate UEs by directly attacking a surrogate model and then transfer the generated UEs to fight against a diverse set of target models~\citep{huang2020unlearnable, fowl2021adversarial}. This may be easily achievable under the label-consistency setting, but may fail badly under the label-agnostic setting. However, even under the label-consistency setting, few works have studied the impact of the surrogate model to the final unlearnable performance. 
To generate effective, and more importantly, transferable UEs under the label-agnostic setting, we need to explore more generic surrogate model selection strategies, especially those that can be tailored to a wider range of unknown target models. Intuitively, the surrogate model should be a classification DNN that contains as many classes as possible so as to facilitate the recognition and protection of billions of images on the Internet. In this paper, we propose to leverage the large-scale Vision-and-Language Pre-trained Models (VLPMs)~\citep{li2019visualbert, radford2021learning, li2021align} like CLIP~\citep{radford2021learning} as the surrogate model. 
Pre-trained on over 400 million text-to-image pairs, CLIP has the power to extract the representation of extremely diverse semantics. Meanwhile, VLPMs are pre-trained with a textual description rather than a one-hot label to align with the image, making them less overfit to the actual class ``labels''. In this work, we leverage the image encoder of CLIP to extract the embeddings of the input images and then use the embeddings to generate more transferable UCs.


We evaluate our UC approach with different backbones and datasets, all in a black-box setting (the protector does not know the attacker's network architecture or the class labels). Cluster-wise unlearnable noise can also prevent unsupervised exploitation against contrastive learning to certain extent, proving its superiority to existing UEs. We also compare UC with existing UE methods against two commercial machine learning platforms: Microsoft {\tt Azure}\footnote{\url{https://portal.azure.com/}} and Baidu {\tt PaddlePaddle}\footnote{\url{https://www.paddlepaddle.org.cn/en/}}. To the best of our knowledge, this is the first physical-world attack to commercial APIs in this line of work. Our main contributions are summarized as follows:

\begin{itemize}
    \item We promote a more generic data protection assumption called \textbf{label-agnostic}, which allows the hackers to exploit the protected dataset differently (in terms of the annotated class labels) as the protector. This opens up a more practical and challenging setting against unauthorized training of machine learning models.
    
    \item We reveal the working mechanism of existing UE generation methods: they all disrupt the distributional uniformity and discrepancy in the deep representation space. 
    
    \item We propose a novel approach called \textbf{Unlearnable Clusters} (UCs) to generate label-agnostic UEs with cluster-wise perturbations without knowing the label information. We also leverage VLPMs like CLIP as the surrogate model to craft more transferable UCs.
    
    \item We empirically verify the effectiveness of our proposed approach with different backbones on different datasets. We also show its effectiveness in protecting private data against commercial machine learning platforms {\tt Azure} and {\tt PaddlePaddle}.
\end{itemize}

\section{Related Work}\label{related}

Unlearnable examples (UEs) can be viewed as one special type of data poisoning attacks~\citep{biggio2012poisoning, biggio2018wild} that aim to make model training fail completely on the poisoned (protected) dataset. UEs should be differentiated from the other two well-known attacks to deep learning models: backdoor attacks~\citep{gu2017badnets, chen2017targeted,liu2020reflection} and adversarial attacks~\citep{szegedy2013intriguing, goodfellow2014explaining}. Backdoor attacks are the other special type of data poisoning attacks that do not impact the model's performance on clean data, which is in sharp contrast to UEs. Adversarial attacks are one type of test-time attacks that evade the model's prediction by adding small imperceptible adversarial noise to the inputs. 

UEs can be generated via a min-min bilevel optimization framework with a surrogate model \citep{huang2020unlearnable}, similar to the generation of strong data poisons via bilevel optimization~\citep{shafahi2018poison, zhu2019transferable, huang2020metapoison, schwarzschild2021just}.
The generated noise is termed Error-Minimizing Noise (EMinN) as it progressively eliminates errors from the training data to trick the target model to believe there is nothing to learn \citep{huang2020unlearnable}. We use EMinN to denote the original UE generation method. 
In addition to EMinN, there are also UE generation methods that utilize adversarial noise, such as Error-Maximizing Noise (EMaxN)~\citep{koh2017understanding}, DeepConfuse~\citep{feng2019learning} and Adversarial Poisoning (AdvPoison)~\citep{fowl2021adversarial}. Recently, Yu et al.~\citep{yu2022availability} unveil a linear-separability property of unlearnable noise and propose the Synthetic Perturbations (SynPer) method to directly synthesize linearly-separable perturbations as effective unlearnable noise.

The original UE method EMinN has a few limitations. First, the generated unlearnable noise can be removed to a large extent by adversarial training \citep{madry2017towards}, although this will also decrease the model's performance by a considerable amount \citep{huang2020unlearnable}. This was later on solved by a recent work published at ICLR 2022~\citep{fu2022robust}.  The idea is to optimize the adversarial training loss in place of the standard training loss to produce more robust error-minimizing noise.  
The other limitation is its transferability to different training schemes, target models (the models to protect against) or datasets. For example, it has been found that unlearnable noise generated in a supervised manner fails to protect the dataset from unsupervised contrastive learning~\citep{he2022indiscriminate}. A unsupervised UE generation method was then proposed to craft UEs unlearnable to unsupervised contrastive learning. However, a very recent work by Ren et al.~\citep{ren2022transferable} demonstrates that, surprisingly, unsupervised UEs cannot protect the dataset from supervised exploitation.
All above UE methods all rely on the ideal label-consistency assumption, i.e., the same (or no) labels for the protected data will be used by both the protectors and hackers. In this paper, we promote a more practical label-agnostic setting where different labels could be used by the hackers for their own purposes.


Besides UEs, strong adversarial attacks have also been proposed to protect personal data from malicious face recognition systems, such as LowKey~\citep{cherepanova2021lowkey} and APF~\citep{zhang2020adversarial}. They differ from UEs by making a normally trained model unable to recognize the protected images, rather than preventing the proper training of any machine learning models on the protected images. In this work, we focus on UEs rather than other data protection techniques which we believe are of independent interest.

\section{Proposed Method}

\paragraph{Threat Model.}
We introduce two parties: the \textbf{protector} and the \textbf{hacker}. The protectors leverage a surrogate model to generate UEs for its private data before publishing it on the Internet. 
For example, online social network companies (or users) could convert their photos to their UE versions before posting them online.
These ``protected" images are then collected, without the protectors' consent, by a hacker into a dataset to train a commercial or malicious model. The protectors' goal is to make the collected dataset unlearnable, i.e., cannot be used for model training, while the hackers' goal is to train accurate models on the unlearnable (protected) dataset. Following prior works \citep{huang2020unlearnable,fu2022robust,liu2021going}, we assume the released dataset is 100\% protected, i.e., all the samples are perturbed to be unlearnable. While this assumption appears to be ideal, if the protection technique is reliable, there is no reason not to employ it to gain more protection and privacy. Therefore, in this work we choose to focus on the unlearnable technique itself rather than changing the setting of the protectors.
Following our label-agnostic setting, we also assume the hackers could exploit the unlearnable dataset with different labels. E.g., a $m$-class dataset could be exploited by the hacker as a $n$-class dataset.

Here, we give an example of such label-agnostic scenario with a online social media company who strives to protect the contents created by all of its users. The company could leverage unlearnable techniques to develop systematic protection scheme against unauthorized data explorers. In this case, we can assume all the images uploaded by the users are protected (by the company). Potential hackers like Clearview AI may crawl the images from the online platform without the users' content into one or a set of datasets for its own purposes. Thus, the collected datasets cannot be guaranteed to have the same labels as their original versions. The protector thus needs to craft more powerful and transferable unlearnable examples to make data unexploitable against different labeling strategies.


\subsection{Problem Formulation}\label{sec:3.1}
We focus on image classification tasks in this paper.
Given a clean $m$-class training dataset $\mathcal{D}_{c}^{m}=\{(\xx_i, \yy_i)\}_{i=1}^{k}$ consisting of $k$ clean training images $\xx \in \X \subset \mathbb{R}^d$ and their labels $\yy \in \Y$, in a standard unlearnable setting \citep{huang2020unlearnable}, the protector trains an $m$-class surrogate model $f_{s}^{m}$ on $\mathcal{D}_{c}^{m}$. The protector can then generate an unlearnable version of the dataset as $\mathcal{D}_{u}^{m} = \{(\xx^{\prime}_i, \yy^{\prime}_i)\}_{i=1}^k$, based on the clean dataset $\mathcal{D}_{c}^{m}$ and the surrogate model $f_{s}^{m}$. The unlearnable images are denoted as $\xx^{\prime} = \xx + \bm{\delta}$ ($\xx \in \mathcal{D}_{c}^{m}$) with the same labels $\yy \in \Y$ as their original versions and $\bm{\delta} \in \Delta \subset \mathbb{R}^d$ are the generated unlearnable noise which is often regularized to be imperceptible.
The unlearnable dataset $\mathcal{D}_{u}^{m}$ is assumed to be the dataset collected by the hackers, and will be exploited to train a commercial or malicious $m$-class target model $f_{t}^{m}$ without the protectors' consent.

\vspace{-10pt}
\paragraph{Label-consistency vs. Label-agnostic.}
The above formulation follows the standard \emph{label-consistency} assumption of previous works \citep{huang2020unlearnable,fu2022robust}, where the hackers collect, annotate and exploit the unlearnable dataset $\mathcal{D}_{u}^{m}$ exactly the same as it was initially released by the protectors. Under a more general and practical \emph{label-agnostic} assumption, the hackers could annotate the collected dataset $\mathcal{D}_{u}^{m}$ differently, e.g., assigning it with different number of classes. 
In this case, the hackers may exploit the dataset as a $n$-class ($n \neq m$) classification dataset $\mathcal{D}_{c}^{n}=\{(\xx^{\prime}_i, \yy^{\prime}_i)\}_{i=1}^{k}$ to train a $n$-class target model $f_{t}^{n}$. Note that the protectors have no knowledge of the target class number $n$ nor the target labels $\yy^{\prime}_i$.
Arguably, the hackers may even exploit the dataset as an object detection dataset rather than a classification dataset. We will explore such a more challenging \emph{task-agnostic} assumption in our future work and focus on the label-agnostic in this work.

\subsection{The Label-agnostic Challenge}

\begin{figure}[t]
\begin{minipage}[b]{0.40\linewidth}
\centering
\includegraphics[width=0.99\textwidth]{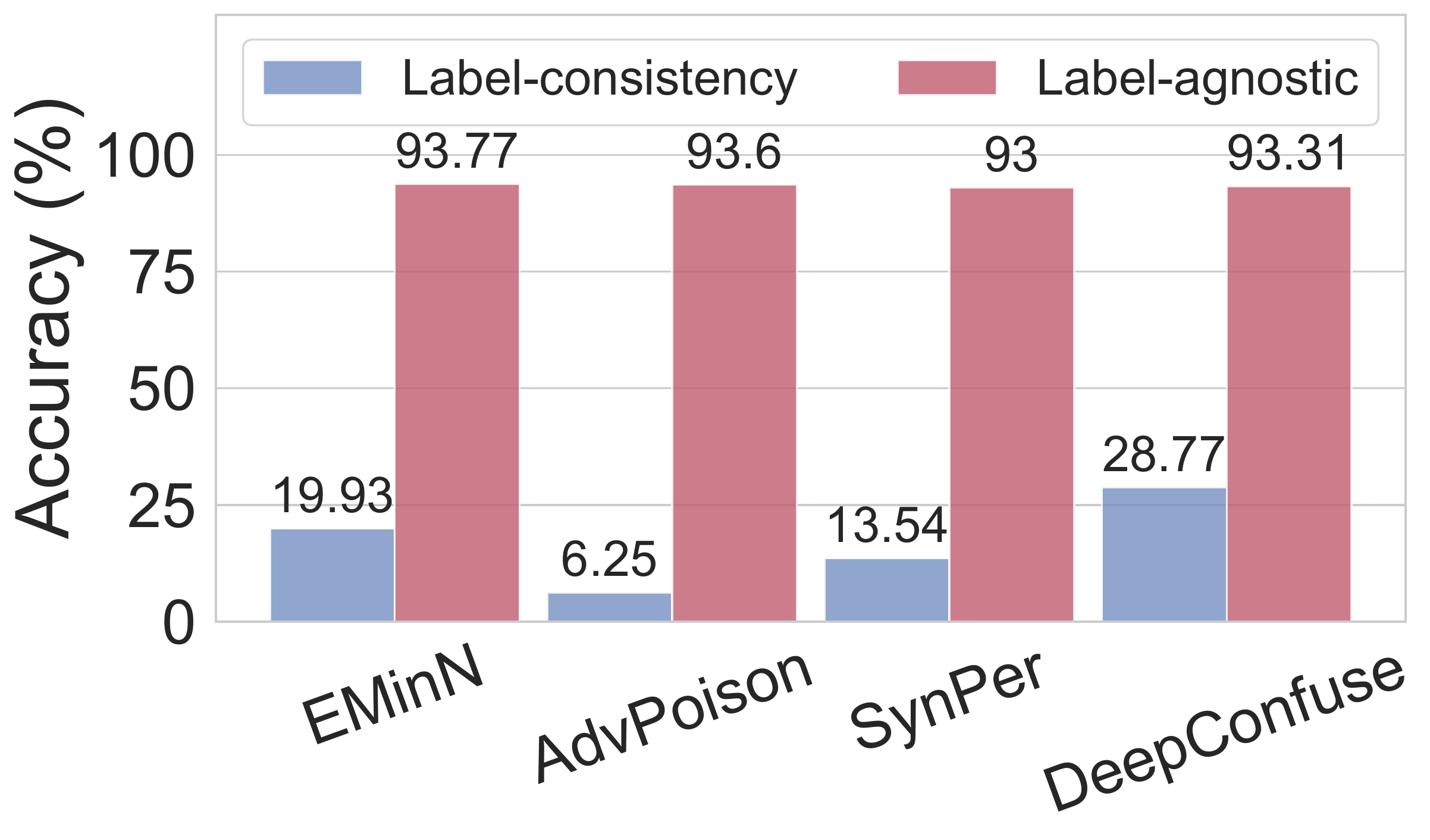}
\centerline{(a)}
\end{minipage}
\begin{minipage}[b]{0.59\linewidth}
\centering
\includegraphics[width=0.99\textwidth]{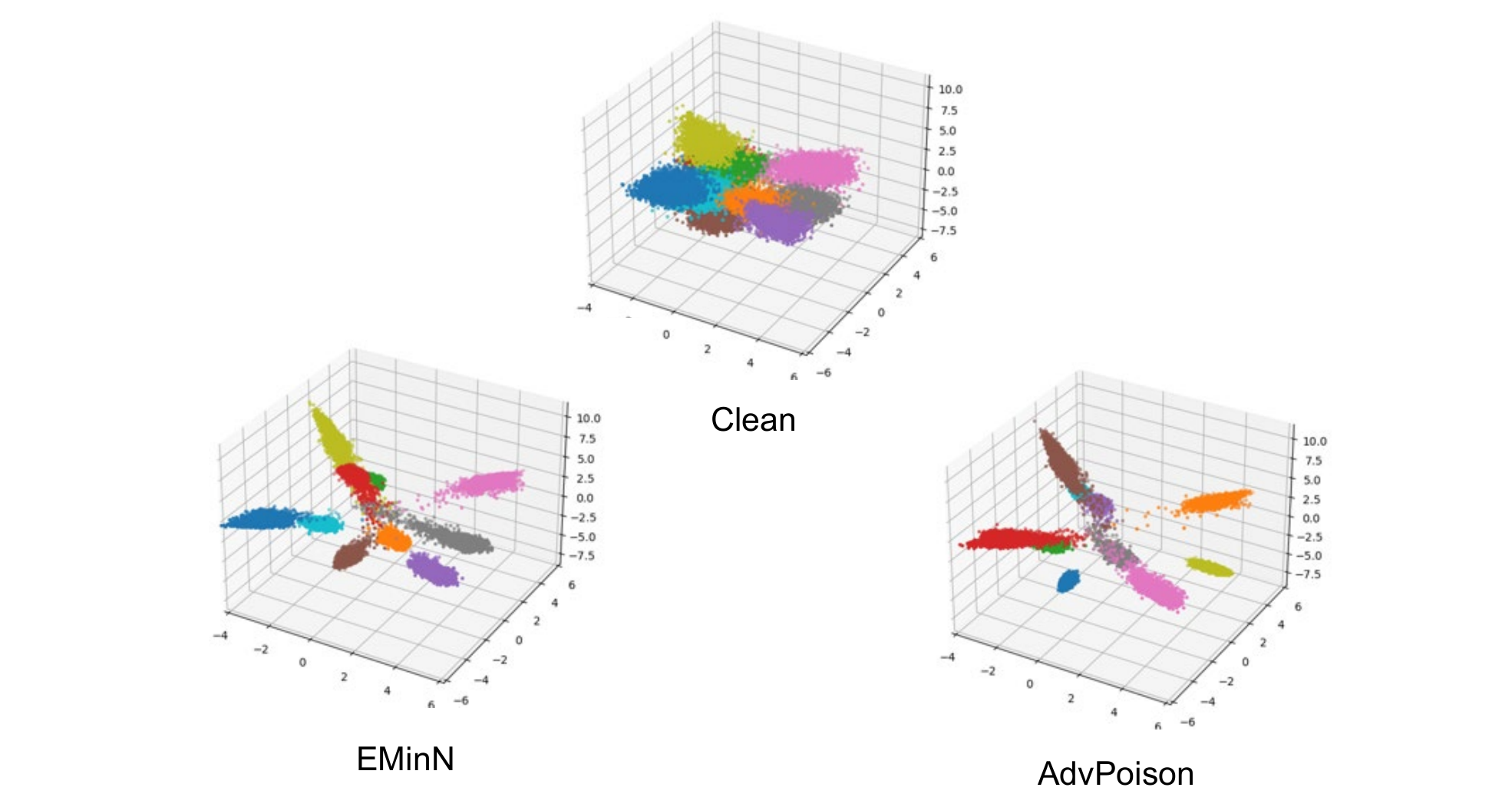}
\centerline{(b)}
\end{minipage}
\vspace{-10pt}
\caption{(a) Current UE methods become ineffective in the label-agnostic setting. (b) A 3D feature visualization of clean CIFAR-10 examples and the UEs derived by EMinN and AdvPoison. Points in the same color denote samples of the same class.}
 \label{fig_2}
\vspace{-10pt}
\end{figure}


\paragraph{Existing methods are not robust to label-agnostic exploitation.}
We test the effectiveness of existing unlearnable methods developed under the label-consistency setting against label-agnostic hackers. Here we consider current unlearnable method including Error-Minimizing Noise (EMinN)~\citep{huang2020unlearnable}, Adversarial Poisoning (AdvPoison)~\citep{fowl2021adversarial}, Synthetic Perturbations (SynPer)~\citep{yu2022availability} and DeepConfuse~\citep{feng2019learning}, on the CIFAR-10 dataset~\citep{krizhevsky2009learning}. The ResNet-18 \citep{he2016deep} models are used for both the surrogate and target models. As shown in Figure ~\ref{fig_2} (a), these methods are extremely effective in preventing the training of machine learning models on the unlearnable dataset with the same labels. However, if the unlearnable dataset is crafted using ImageNet surrogate model with the predicted ImageNet labels (i.e., labels predicted by the surrogate model), it fails to prevent the model training with the original CIFAR-10 labels. This indicates one unique challenge of the label-agnostic setting: \emph{unlearnable noises generated to prevent one set of labels are not transferable to preventing other labeling strategies}.

\begin{figure*}[t]
  \centering
  \includegraphics[width=\linewidth]{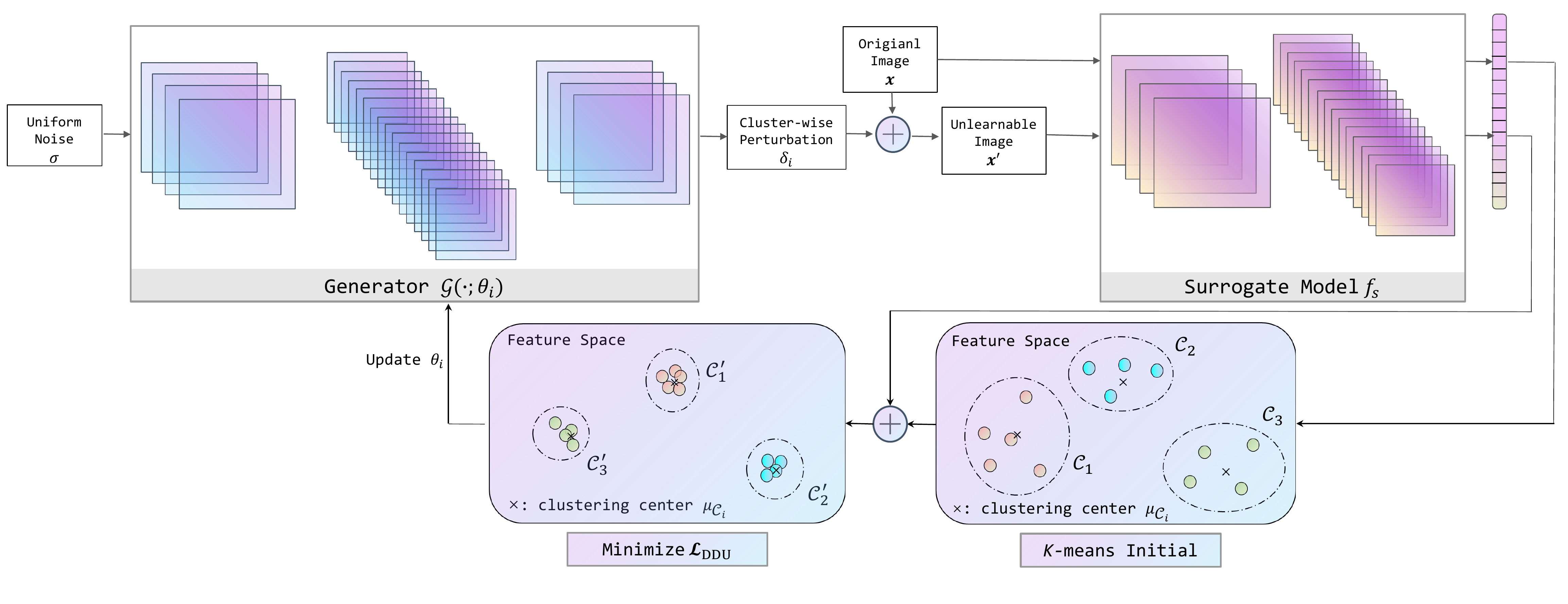}
  \vspace{-25pt}
  \caption{The Unlearnable Clusters pipeline. The entire dataset is divided into $p$ clusters via K-means clustering, where each cluster corresponds to a certain generator with parameters $\theta_{i}$ and a cluster-wise perturbation $\bm{\delta}_i$.}
  \label{fig_framework}
\vspace{-15pt}
\end{figure*}

\paragraph{The working mechanism of existing UEs under the label-consistency setting.}
Here, we investigate the representations learned by the target model on clean vs. unlearnable examples, aiming to gain more understanding of the unlearnable mechanism.
In Figure~\ref{fig_2} (b), we visualize the 3-dimensional PCA~\citep{wold1987principal} projections of the original representations learned by the ResNet-18 target model for a) clean CIFAR-10 training samples, b) unlearnable CIFAR-10 examples crafted by EMinN method, and 3) unlearnable (poisoned) CIFAR-10 examples crafted by AdvPoison. It shows in Figure~\ref{fig_2} (b) that the unlearnable examples crafted by EMinN and AdvPoison tend to significantly reduce the variance at certain dimensions. There are also classes that collapse into smaller clusters, like the green class. This indicates that the noise disrupts the \emph{distributional discrepancy} in the representation space to make the data ``unlearnable". The other key observation is that the noise greatly shifts the points away from the normal data manifold, causing an unnecessary spread over a certain direction. This indicates that the noise also breaks the \emph{distributional uniformity} of the data. Overall, it is evident the unlearnable noise crafted by EMinN and AdvPoison cripples the learning process by distorting both the discrepancy and uniformity of the data distribution in the deep representation space.

\paragraph{Unlearnable examples can overfit to the labels.}
A closer look at the visualizations in Figure~\ref{fig_2} (b), one may notice that the unlearning effects occur only within the classes. I.e., the UEs have overfitted to the class labels. This is somewhat not surprising as the unlearnable noises are generated via a supervised loss function (i.e., cross-entropy) defined by the labels. The noise are thus optimized to thwart the most predictive information to the class labels. However, this causes the overfitting problem and fails to work if the labels are changed. Intuitively, if we could remove the dependency on the class labels and turn to exploit the clusters that naturally arise during the learning process, we could make the unlearnable noise more robust to different annotations.

\subsection{Unlearnable Clusters (UCs)}

\paragraph{Overview.}
Motivated by the above observations, in this work we propose to generate UEs by exploiting the clusters learned by a surrogate model and making the clusters unlearnable instead of the labeled classes.
We term this approach as \textbf{Unlearnable Clusters (UCs)} and illustrate its workflow in Figure~\ref{fig_framework}. The key components of UC are one generator model $\mathcal{G}$ and one surrogate model $f_s$. 
At a high level, UC first employs a surrogate model $f_s$ to extract the representations $\EE$ of all samples in the clean dataset $\mathcal{D}_{c}$. It then utilizes the K-means~\citep{selim1984k} clustering method to derive $p$ clusters from the representations $\EE$. Subsequently, for each cluster, it generates a cluster-wise perturbation $\bm{\delta}_i$ using the generator $\mathcal{G}$. The noise will be generated and applied to craft the UE for each sample in $\mathcal{D}_{c}$, with samples belonging to the same cluster are added with the same cluster-wise noise $\bm{\delta}_i$. UEs crafted in this manner 
can prevent the target model from learning meaningful clusters rather than class predictions, thus is more general to different types of label exploitations. Next, we will introduce the details of UCs.

\vspace{-10pt}
\paragraph{Cluster-wise Perturbations.}
In our UC framework, one encoder-decoder~\citep{poursaeed2018generative} generator network is used to generate the cluster-wise perturbations, with each generator will be reinitialized for one cluster. 
As such, we need to extract the clusters first. Here, we leverage the most classic clustering method K-means~\citep{selim1984k} to detect clusters from the deep representations. Particularly, the clean dataset $\mathcal{D}_{c}$ is fed into the surrogate model $f_s$ to extract the representation matrix before the classification layer $\EE = [\ee_1, \cdots, \ee_k]$. K-means clustering is then applied on the representation matrix to  detect $p$ number of clusters $\mathcal{C}=\{ \mathcal{C}_1, \cdots, \mathcal{C}_p \}$, where $\mathcal{C}_i= \{ \xx_{ij} \}^{\tau(i)}_{j=1} = \{ \xx_{i1}, \cdots, \xx_{i \tau(i)} \}$ and $\sum_{i=1}^{p} \tau(i)=k$. The corresponding centers for the clusters are $\mu_{\mathcal{C}}=\{ \mu_{\mathcal{C}_1}, \cdots, \mu_{\mathcal{C}_p} \}$. 

With the detected clusters $\mathcal{C}$, we can now propose the following method to generate the unlearnable noise for each cluster.
Intuitively, for cluster $\mathcal{C}_i$, we hope the unlearnable noise $\bm{\delta}_i$ could move all samples in the cluster to a \emph{wrong} cluster center, so as to force the model to forget the correct clusters. This is done via the following minimization framework:
\begin{equation}
\begin{aligned}
\label{eq_1}
\theta_{i} &= \argmin_{\theta_{i}} \mathcal{L}_{\rm DDU} (\mathcal{C}_{i}, g(\mu_{\mathcal{C}_i}), \theta_{i}) \\
&= \argmin_{\theta_{i}} \sum_{\xx_{ij} \in \mathcal{C}_{i}} d( f_s(\xx_{ij} + \mathcal{G}(\sigma; \theta_{i})), g(\mu_{\mathcal{C}_i}) ),
\end{aligned}
\end{equation}
where, $\mathcal{L}_{\rm DDU}$ is our proposed Disrupting Discrepancy and Uniformity (DDU) loss that defines the distance ($d(\cdot)$) of samples in $\mathcal{C}_i$ to a permuted (wrong) cluster center by a permutation function $g(\mu_{\mathcal{C}_i})$; $\theta_i$ are the parameters of generator network $\mathcal{G}$; $\mathcal{G}(\sigma; \theta_{i}))$ generates the unlearnable noise for all samples in $\mathcal{C}_i$ (i.e., $\xx_{ij} \in \mathcal{C}_i$).
Please note that the above problem needs to be solved for $p$ times to obtain the cluster-wise unlearnable noise for all  $p$ clusters, and for each cluster, the generator $\mathcal{G}$ is reinitialized with new parameters $\theta_{i}$.
The complete procedure is described in Algorithm~\ref{algorithm}.


\begin{algorithm}[ht!]
\caption{Unlearnable Cluster Generation}
\label{algorithm}
\begin{algorithmic}[1]
\State {\bfseries Input:} surrogate model $f_s$, distance metric $d$, uniform noise $\sigma$, number of clusters $p$, random permutation $g$, $L_{\infty}$-norm restriction $\epsilon$, clean images $\xx \in D_{c}$, initialized generator $\mathcal{G}$ with parameters $\theta$
\State {\bfseries Output:} cluster-wise perturbations $\bm{\delta}=\{\bm{\delta}_1, \cdots, \bm{\delta}_p \}$ 
\State feature matrix $\EE = f_s(\xx)$
\State clusters and cluster centers $\{ \mathcal{C}, \mu_{\mathcal{C}} \} =$ K-means$(\EE,p)$

\For{$i$ {\bfseries in} $1 \cdots p$}
    \State Initialize $\theta_i$
    \State $\bm{\delta}_i$ = $\mathcal{G}(\sigma; \theta_i)$
    \State $\bm{\delta}_i$ = Clamp($\bm{\delta}_i, -\epsilon, \epsilon$)
    \For{$\xx_{ij}$ {\bfseries in} $\mathcal{C}_{i}$}
        \State $\xx_{ij}^{\prime} =$ Clamp($\xx_{ij}+\bm{\delta}_i, 0, 1$)
        \State $\theta_{i} \leftarrow$ Optimize($\xx_{ij}^{\prime}, f_s, g(\mu_{\mathcal{C}_i}), d$)
    \EndFor
    \State $\bm{\delta}_i$ = $\mathcal{G}(\sigma; \theta_i)$
    \State $\bm{\delta}_i$ = Clamp($\bm{\delta}_i, -\epsilon, \epsilon$)
\EndFor

\end{algorithmic}
\end{algorithm}

\paragraph{CLIP Surrogate Model.}
How to choose a surrogate model remains to be an independent challenge for generating effective cluster-wise unlearnable noise. As shown in prior works, it plays a central role in facilitating the transferability of the generated UEs to different datasets or target models~\citep{huang2020unlearnable}. 
In the traditional label-consistency setting, the surrogate model can be a model that directly trained on the original (unprotected) dataset, which may of a different (and plausibly a better or more complex) model architecture. It could also be a model that trained on a larger dataset with more classes, e.g., ImageNet-trained models ~\citep{huang2020unlearnable,fowl2021adversarial}. We thus adopt an ImageNet-pretrained ResNet-50 as the default surrogate model of our UC. 

Analogous to the classification surrogate models used for generating the traditional UEs, the ideal surrogate models for unlearnable clusters could be those powerful feature extractors that could lead to accurate detection of clusters from an image dataset.
We thus propose to also leverage one large-scale vision-and-language pre-trained model (VLPM)~\citep{li2019visualbert,li2021align} CLIP~\citep{radford2021learning} as our surrogate model. Pre-trained on over $400$ million text-to-image pairs, CLIP has the power to extract the representation of extremely diverse semantics. Moreover, CLIP was pre-trained with a textual description rather than a one-hot label to align with the image, thus overfitting less to the actual class labels. 
Concretely, we employ the image encoder of CLIP to extract the feature matrix for the clean dataset, which is then used to compute the clusters and cluster centers. We denote the version of UC equipped with the CLIP surrogate model as UC-CLIP.

\section{Experiments}\label{sec_exp}

In this section, we evaluate our UCs methods on different datasets against different target models, which is to simulate as many unknown cases as possible. We also examine the robustness of UCs against several advanced defenses. Finally, we demonstrate its effectiveness in attacking commercial machine learning platforms {\tt Azure} and {\tt PaddlePaddle}.

\subsection{Experimental Settings}\label{sec_exp_1}

\paragraph{Datasets and Models.}
We conduct our study on $6$ high-resolution and industrial-scale vision datasets to simulate as diverse real-world applications as possible, including Pets~\citep{parkhi2012cats}, Cars~\citep{krause20133d}, Flowers~\citep{nilsback2008automated}, Food~\citep{bossard2014food}, SUN397~\citep{xiao2010sun} and ImageNet~\citep{russakovsky2015imagenet}. For ImageNet, we only use its first $100$ classes which is denoted as ImageNet$^\star$. For surrogate models, we consider ResNet-50 trained on ImageNet-1k as the default, unless otherwise explicitly stated. For target models, we employ randomly initialized ResNet-18~\citep{he2016deep}, EfficientNet-B1~\citep{tan2019efficientnet} and RegNetX-1.6GF~\citep{radosavovic2020designing}. Notice that we train the target models with data augmentations (resizing, random crop, random horizontal flip and normalization).

For each $\bm{\delta}_i$, we repeated $p$ times to train the generator $\mathcal{G}$ for $10$ epochs for entire ImageNet$^\star$ and $50$ epochs for other entire datasets. For random permutation $g(\cdot)$, we simply chose $i \rightarrow i+1$ to build a closed loop. We consider $L_{\infty}$-norm restriction in this work, i.e., $\|\bm\delta\|_{\infty} < \epsilon=16/255$. The number of clusters $p$ is set to $10$, with an analysis is provided in Section~\ref{sec_exp_5}.

\paragraph{Baselines.}
We compare our UC and UC-CLIP with 5 baseline methods including DeepConfuse~\citep{feng2019learning}, Synthetic Perturbations (SynPer)~\citep{yu2022availability}, Error-minimizing Noise (EMinN)~\citep{huang2020unlearnable}, Error-maximizing Noise(EMaxN)~\citep{koh2017understanding}, and Adversarial Poisoning (AdvPoison)~\citep{fowl2021adversarial}.

\paragraph{Label-agnostic Setup.} Please note that we conduct all of our experiments under the proposed label-agnostic setting. The UCs (and the UEs they serve) are all generated with the predicted labels by the surrogate models. The predicted labels may overlap with the ground truth labels to some extent, but are highly inconsistent with the original labels. We report the test accuracy of the target models on the respective clean test sets.

\subsection{Main Results}\label{sec_exp_2}

\begin{table*}[t]
\centering
\caption{The test accuracy (\%) of different target models trained on the unlearnable datasets generated by our UC/UC-CLIP and the 5 baseline methods, under the label-agnostic setting. The top-2 best results are highlighted in \textbf{bold}.}
\label{tab_1}
\begin{sc}
\renewcommand\arraystretch{1.7}
\resizebox{1.0\linewidth}{!}{
\setlength{\tabcolsep}{0.2mm}{
\begin{tabular}{c |C{1cm}ccccc|C{1cm}ccccc|C{1cm}ccccc} 
\toprule[1pt]
 \rowcolor{white} &  \multicolumn{6}{|c}{{\large ResNet-18}} & \multicolumn{6}{|c}{{\large EfficientNet-B1}} & \multicolumn{6}{|c}{{\large RegNetX-1.6GF} } \\ \cline{2-19}
 {\large Methods} & Pets & Cars & Flowers & Food & SUN397 & ImageNet$^\star$ & Pets & Cars & Flowers & Food & SUN397 & ImageNet$^\star$ & Pets & Cars & Flowers & Food & SUN397 & ImageNet$^\star$\\
\hline
{\large Clean}                      & $62.31$ & $67.18$ & $67.18$ & $78.97$ & $43.08$ & $77.76$ & $48.68$ & $72.33$ & $52.46$ & $80.29$ & $42.84$ & $78.04$ & $44.86$ & $63.84$ & $52.69$ & $84.02$ & $43.27$ & $80.78$ \\
{\large SynPer}    & $52.60$ & $53.50$ & $52.74$ & $74.80$ & $38.26$ & $74.69$ & $28.02$ & $58.34$ & $42.93$ & $74.99$ & $35.92$ & $72.94$ & $34.51$ & $45.54$ & $47.16$ & $77.65$ & $37.78$ & $60.38$ \\
{\large EMaxN}     & $54.70$ & $52.95$ & $51.70$ & $73.77$ & $37.57$ & $73.82$ & $33.71$ & $55.64$ & $42.66$ & $74.40$ & $37.30$ & $73.72$ & $34.26$ & $43.40$ & $46.25$ & $78.76$ & $37.82$ & $76.72$ \\
{\large EMinN}     & $52.96$ & $54.43$ & $50.58$ & $75.47$ & $38.48$ & $74.20$ & $36.88$ & $54.23$ & $44.06$ & $75.54$ & $37.20$ & $72.20$ & $37.04$ & $39.67$ & $47.34$ & $79.43$ & $36.82$ & $74.86$ \\
{\large AdvPoison}      & $50.86$ & $51.91$ & $50.64$ & $75.07$ & $38.51$ & $73.76$ & $37.99$ & $50.08$ & $\bf{41.65}$ & $74.88$ & $36.44$ & $72.54$ & $34.29$ & $46.06$ & $47.41$ & $78.64$ & $36.42$ & $76.32$ \\
{\large DeepConfuse}                & $53.72$ & $51.11$ & $50.94$ & $73.13$ & $34.41$ & $55.12$ & $35.54$ & $47.15$ & $43.28$ & $72.91$ & $35.22$ & $45.74$ & $33.71$ & $41.15$ & $46.01$ & $77.26$ & $33.52$ & $49.88$ \\
\hline
{\large \textbf{UC (Ours)}}         & $\bf{12.21}$  & $\bf{33.57}$ & $\bf{35.55}$ & $\bf{55.29}$ & $\bf{20.38}$ & $\bf{54.80}$ & $\bf{17.06}$  & $\bf{13.92}$ & $42.28$ & $\bf{53.45}$ & $\bf{22.97}$ & $\bf{32.30}$ & $\bf{4.28}$  & $\bf{29.46}$ & $\bf{33.79}$ & $\bf{64.48}$ & $\bf{22.28}$ & $\bf{56.10}$ \\
{\large \textbf{UC-CLIP (Ours)}}& $\bf{4.69}$  & $\bf{4.74}$ & $\bf{10.07}$  & $\bf{19.07}$ & $ \bf{3.89}$ & $\bf{39.78}$ & $\bf{6.49}$  & $\bf{15.33}$ & $\bf{14.13}$  & $\bf{17.44}$ & $ \bf{12.95}$ & $\bf{31.82}$ & $\bf{3.87}$  & $\bf{4.18}$ & $\bf{8.12}$  & $\bf{26.76}$ & $ \bf{6.04}$ & $\bf{41.66}$ \\
\bottomrule[1pt]
\end{tabular}}}
\end{sc}
\vspace{-10pt}
\end{table*}

\paragraph{Effectiveness against different target models.}
We first compare our UC and UC-CLIP with the 5 baselines against different target models. Table~\ref{tab_1} shows the results against ResNet-18, EfficientNet-B1, and RegNetX-1.6GF. We have the following main findings: \textbf{(1)} Our methods outperform the baselines by a huge margin consistently across different datasets and target models. This demonstrates the superiority of our methods over the baselines. \textbf{(2)} Our UC-CLIP achieves a better performance than UC, and in most of the cases, by a considerable margin. This proves the great potential of using CLIP as the surrogate model to protect person data from unauthorized exploitations.


\begin{figure}[t]
\begin{minipage}[b]{0.49\linewidth}
\centering
\includegraphics[width=0.99\textwidth]{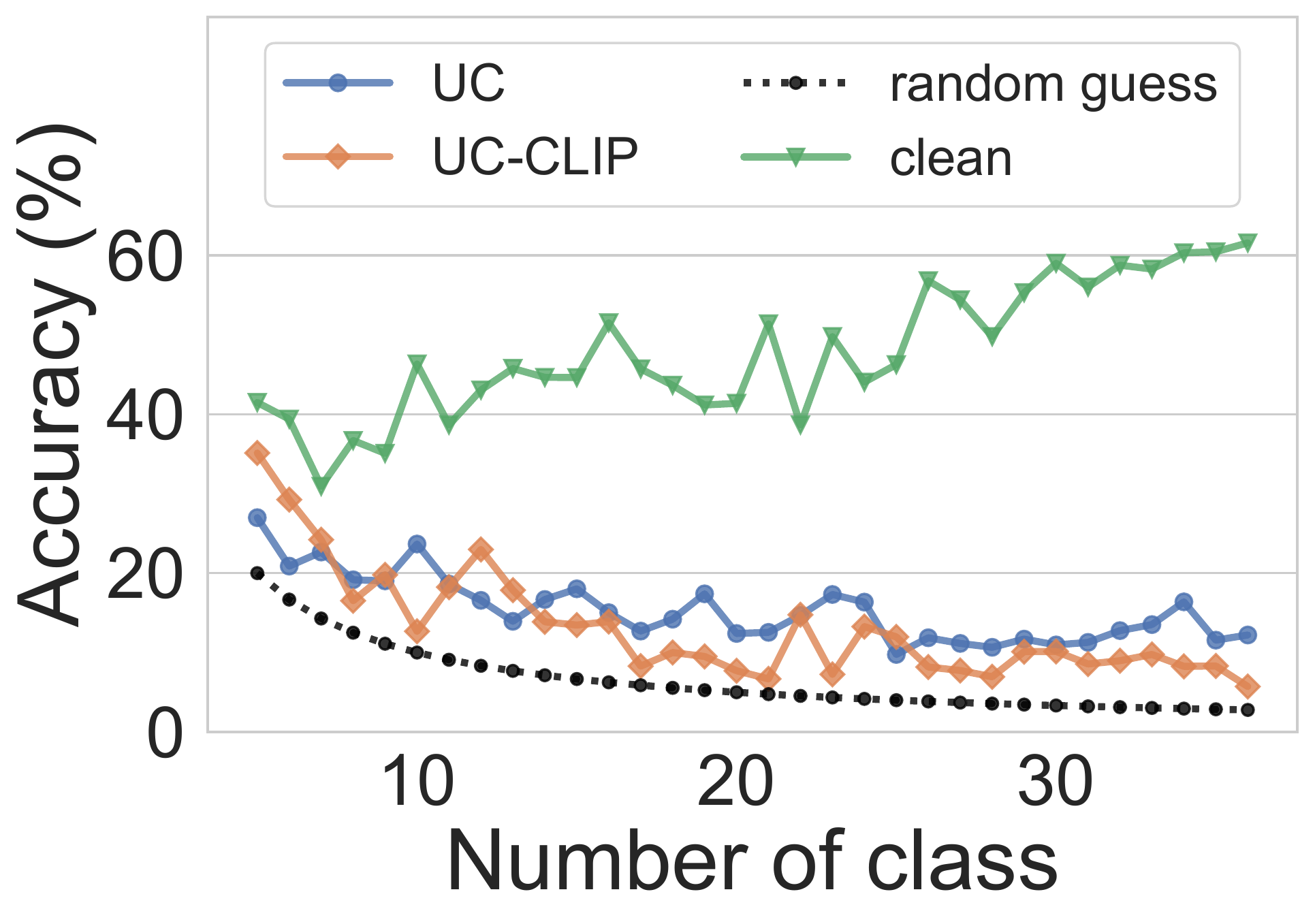}
\centerline{(a) Different labelings}
\end{minipage}
\begin{minipage}[b]{0.49\linewidth}
\centering
\includegraphics[width=0.99\textwidth]{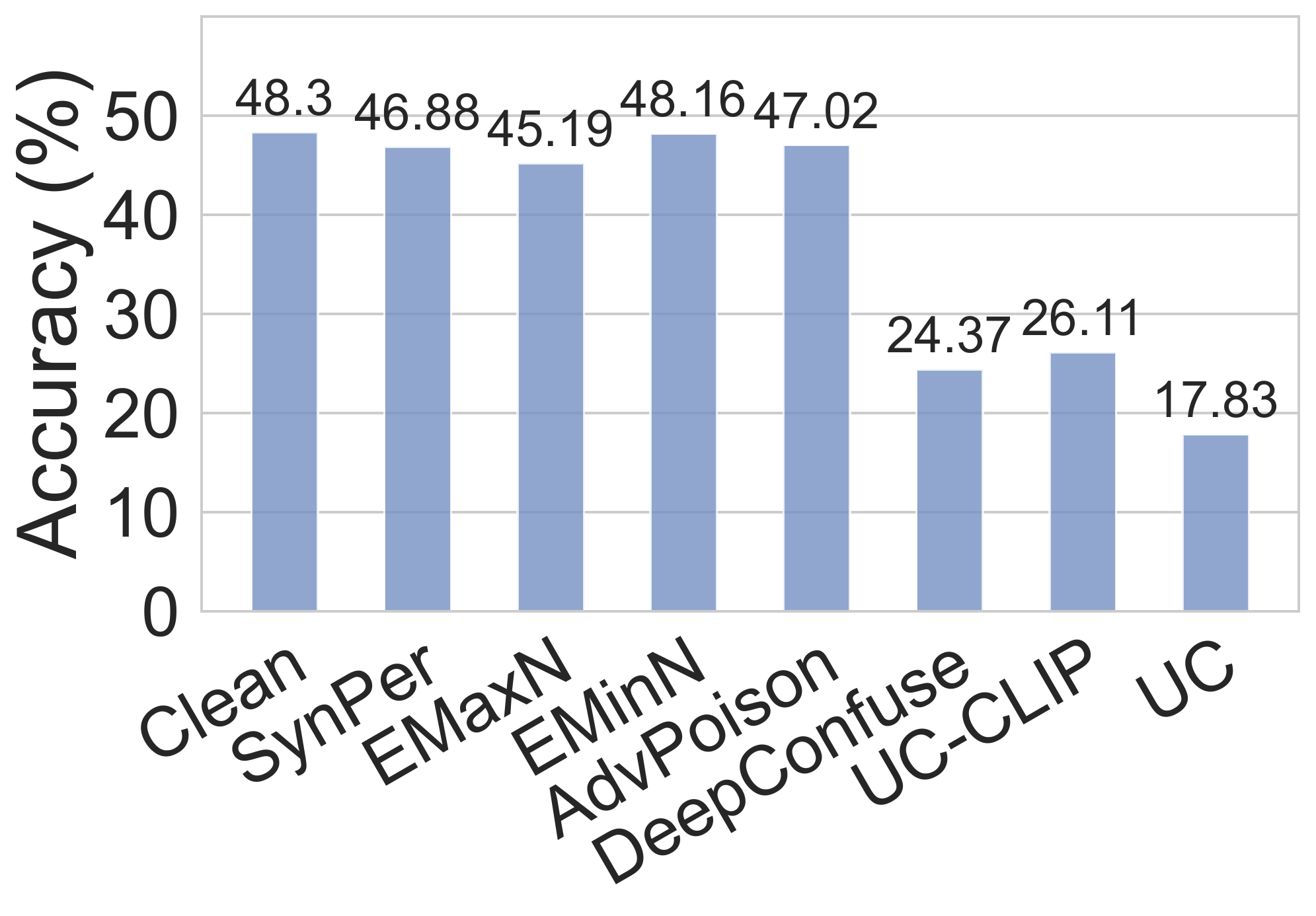}
\centerline{(b) Unsupervised exploitation}
\end{minipage}
\caption{(a) The accuracy of ResNet-18 target models trained on the unlearnable Pets dataset but with its labels were re-labeled by the hacker into 5 to 35 classes. (b) Comparison of our approach with the baselines on Pets dataset against ResNet-18 target model trained via self-supervised SimCLR.}
 \label{fig_4}
\vspace{-15pt}
\end{figure}

\paragraph{Effectiveness Against Different Labelings.}
An even more challenging label-agnostic setting is that the hacker may exploit the unlearnable dataset using different labeling strategies instead of one. So, a natural question is that what if the number of labeled classes of the unlearnable dataset is less than our cluster number $p=10$? Here, we take the 37-class Pets dataset as an example and explore the impact if the hacker re-labels the unlearnable version of the dataset as a 5 to 36 class dataset. One possible labeling strategy is that the hacker first extracts the embeddings of the original text labels using the BERT model~\citep{devlin2018bert}, and then clusters the embeddings into 5-37 classes using K-means, so as to construct a mapping from the old labels to the new labels. As shown in Figure~\ref{fig_4} (a), both our UC and UC-CLIP can bring the test accuracy of the target model down to a level that is close the random guess (the black curve).
This verifies that our methods can craft more generic UEs against the most severe label-agnostic exploitations.

\vspace{-5pt}
\paragraph{Robustness to Unsupervised Exploitation.}
We also compare our methods with the baselines under an unsupervised contrastive learning setting against SimCLR~\citep{chen2020simple}. Although our UC methods are not specifically designed for this unsupervised setting, Figure~\ref{fig_4} (b) shows that cluster-wise unlearnable noise can also prevent unsupervised exploitation against SimCLR.


\subsection{Preventing Commercial Platforms}\label{sec_exp_3}

\begin{table}[t]
\caption{The test accuracy (\%) of models trained by {\tt Azure} and {\tt PaddlePaddle} platforms on unlearnable Cars dataset crafted by different methods. The training configuration on the platform was set to ``fastest training".}
\vspace{-15pt}
\label{tab_2}
\begin{center}
\begin{small}
\begin{sc}
\begin{tabular}{ccc}
\toprule
Methods & {\tt Azure} & {\tt PaddlePaddle} \\
\hline
Clean & $48.45$ & $83.74$ \\
SynPer & $42.38$ & $47.59$ \\ 
EMaxN & $42.83$ & $42.99$ \\
EMinN & $44.06$ & $44.40$ \\
AdvPoison & $43.97$ & $43.38$ \\
DeepConfuse & $39.47$ & $41.88$ \\
\hline
UC (RN50) & $\bf{36.40}$ & $\bf{30.96}$ \\
UC-CLIP (RN50) & $\bf{26.97}$ & $\bf{25.79}$ \\
UC-CLIP (ViTB32) & $\bf{22.47}$ & $\bf{11.49}$ \\
\bottomrule
\end{tabular}
\end{sc}
\end{small}
\end{center}
\vspace{-15pt}
\end{table}

Here, we apply our UC methods to prevent two commercial machine learning platforms: Microsoft {\tt Azure} and Baidu {\tt PaddlePaddle}. On both platforms, the training details are agnostic to us, including the model architecture, learning rate, batch size, epoch, data augmentation, splitting of the validation set, etc. Considering that ViT may be used on commercial platforms due to its recent popularity, we upgrade our UC-CLIP method by replacing the ResNet-50 (RN50) surrogate model by a ViT-B-32 (ViTB32) surrogate model. The results are reported in Table~\ref{tab_2}, which are consistent with that in Table~\ref{tab_1}. I.e., both of our methods can protect the data uploaded to the two platforms against their training algorithms. Unsurprisingly, the ViTB32-powered UC-CLIP method achieves the best protection performance by causing the lowest test accuracy. This suggests the effectiveness of our methods even against commercial platforms.




\subsection{Resistance to Potential Defenses}\label{sec_exp_4}

In this section, we test the robustness of our UC methods to several augmentation based defenses, including Mixup~\citep{zhang2017mixup}, Gaussian smoothing, Cutmix~\citep{yun2019cutmix} and Cutout~\citep{cubuk2018autoaugment}.
As can be observed in Table~\ref{tab_3}, the 4 data augmentation defenses have minimum impact on our UC and UC-CLIP methods. Particularly, Gaussian smoothing appears to be the most effective defense, but the accuracy is still below 25\%.

\begin{table}[t]
\centering
\caption{The test accuracy (\%) of ResNet-18 trained using different defenses against our methods on Pets dataset.}
\vspace{-5pt}
\label{tab_3}
\begin{sc}
\resizebox{1.0\linewidth}{!}{
\begin{tabular}{lcccccc} 
\toprule[1pt]
Methods& No Defense & Mixup & Gaussian & Cutmix & Cutout\\
\midrule
UC      & $12.21$  & $14.34$ & $24.26$ & $14.50$ & $12.35$ \\
UC-CLIP & $4.69$  & $11.96$ & $18.59$ & $6.21$ & $12.29$ \\
\bottomrule[1pt]
\end{tabular}}
\end{sc}
\vspace{-10pt}
\end{table}

\subsection{Ablation Study}\label{sec_exp_5}

\vspace{-5pt}
\begin{figure}[ht]
\begin{minipage}[b]{0.49\linewidth}
\centering
\includegraphics[width=\textwidth]{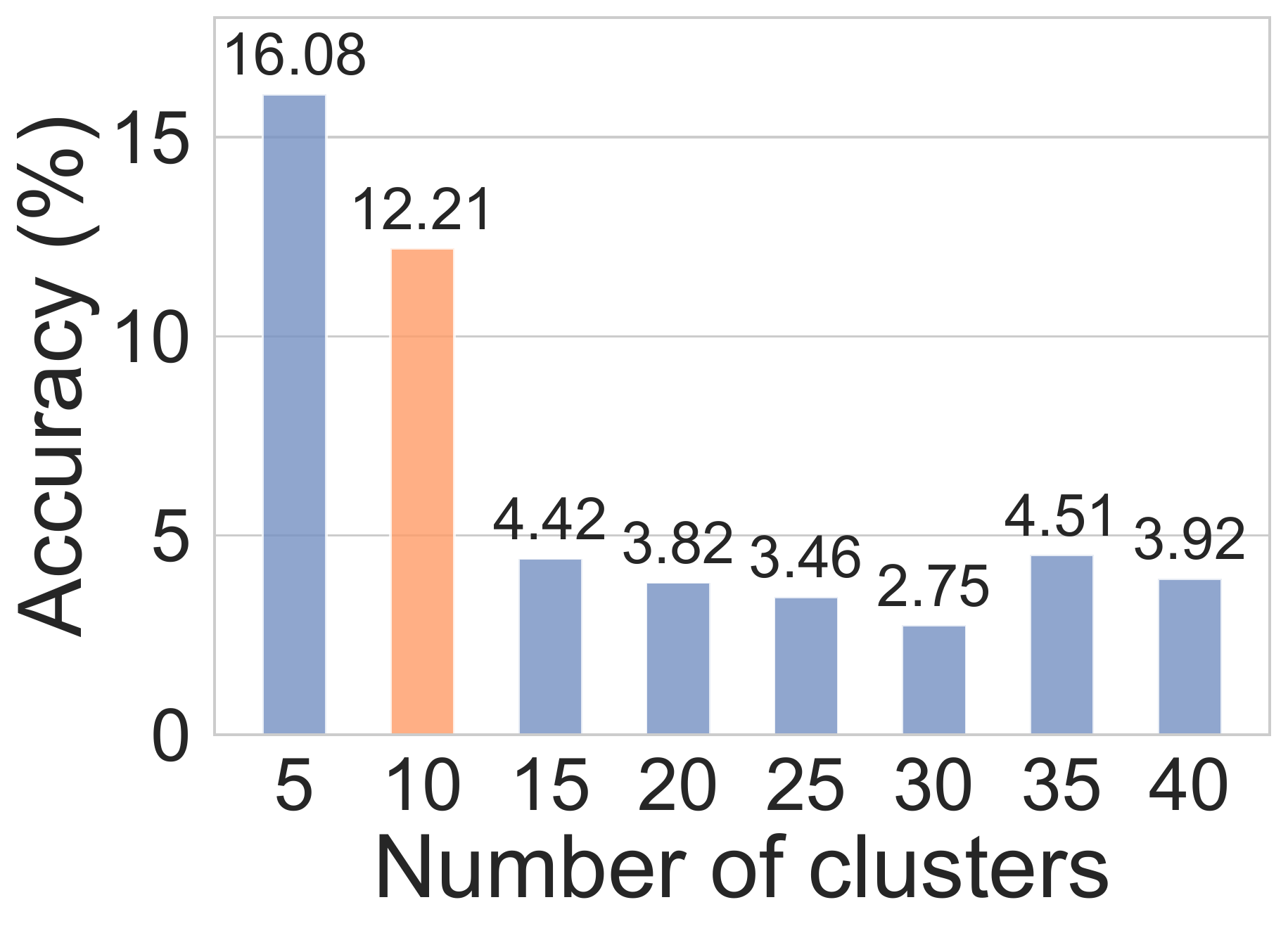}
\vspace{-5pt}
\centerline{(a) Effect of $p$ on UC}
\end{minipage}
\begin{minipage}[b]{0.49\linewidth}
\centering
\includegraphics[width=\textwidth]{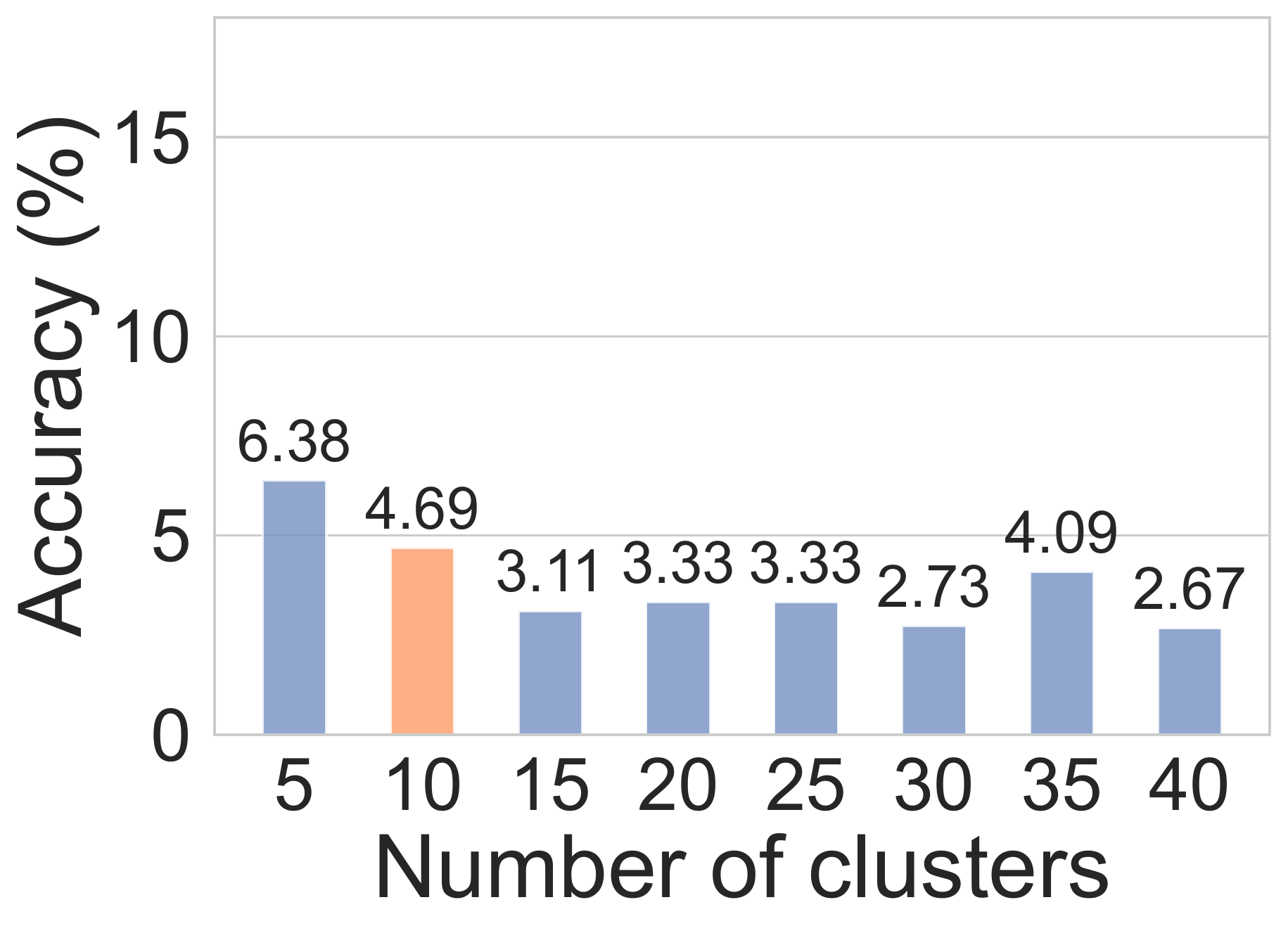}
\centerline{(b) Effect of $p$ on UC-CLIP}
\end{minipage}
\caption{Analyzing the effect of cluster number $p$ on Pets dataset.}
 \label{fig_5}
\vspace{-5pt}
\end{figure}

Here, we analyze the sensitivity of our methods to the number of clusters $p$, which has been set to $p = 10$ as a default. We take the 37-class Pets dataset as an example and evaluate our UC and UC-CLIP method under different values of $p \in [5,40]$. As shown in Figure~\ref{fig_5}, our methods are quite stable to varying hyperparameter $p$ for $p \geq 10$. 
This indicates that, as long as the clusters can cover most of the concepts in a dataset, the generated unlearnable noise can effectively prevent the model from learning the real content from the dataset. As the number of clusters increases, the noise tends to become more effective, although there is a slight variation at $35$.
Note that, even in the worst case at $p=5$, our methods still outperform the baselines.

\subsection{Mixture of Clean and Unlearnable Data}\label{sec_exp_6}
\vspace{-5pt}

\begin{figure}[ht]
\begin{minipage}[b]{0.49\linewidth}
\centering
\includegraphics[width=\textwidth]{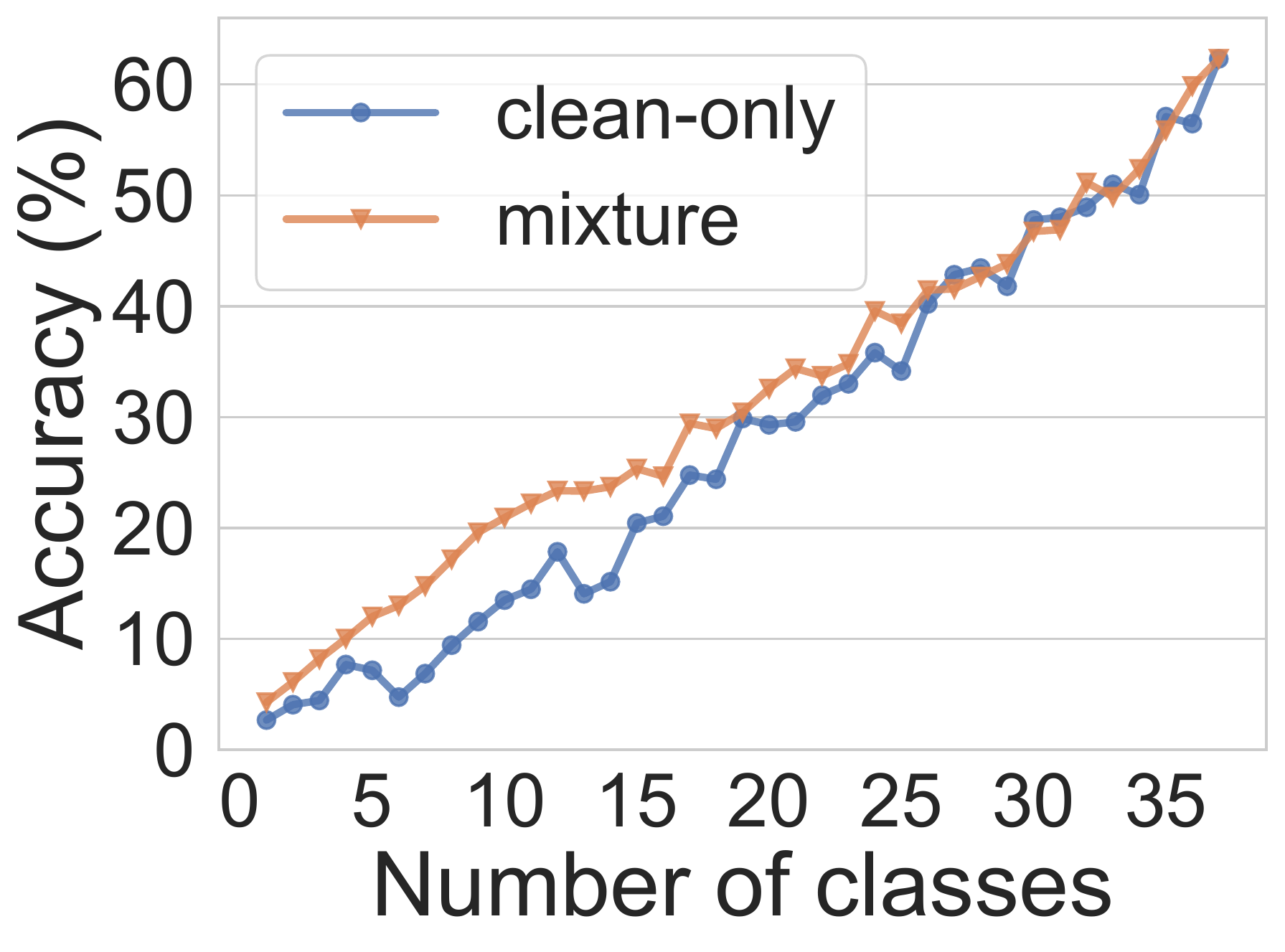}

\centerline{(a) Mixture vs. Clean-only}
\end{minipage}
\begin{minipage}[b]{0.49\linewidth}
\centering
\includegraphics[width=\textwidth]{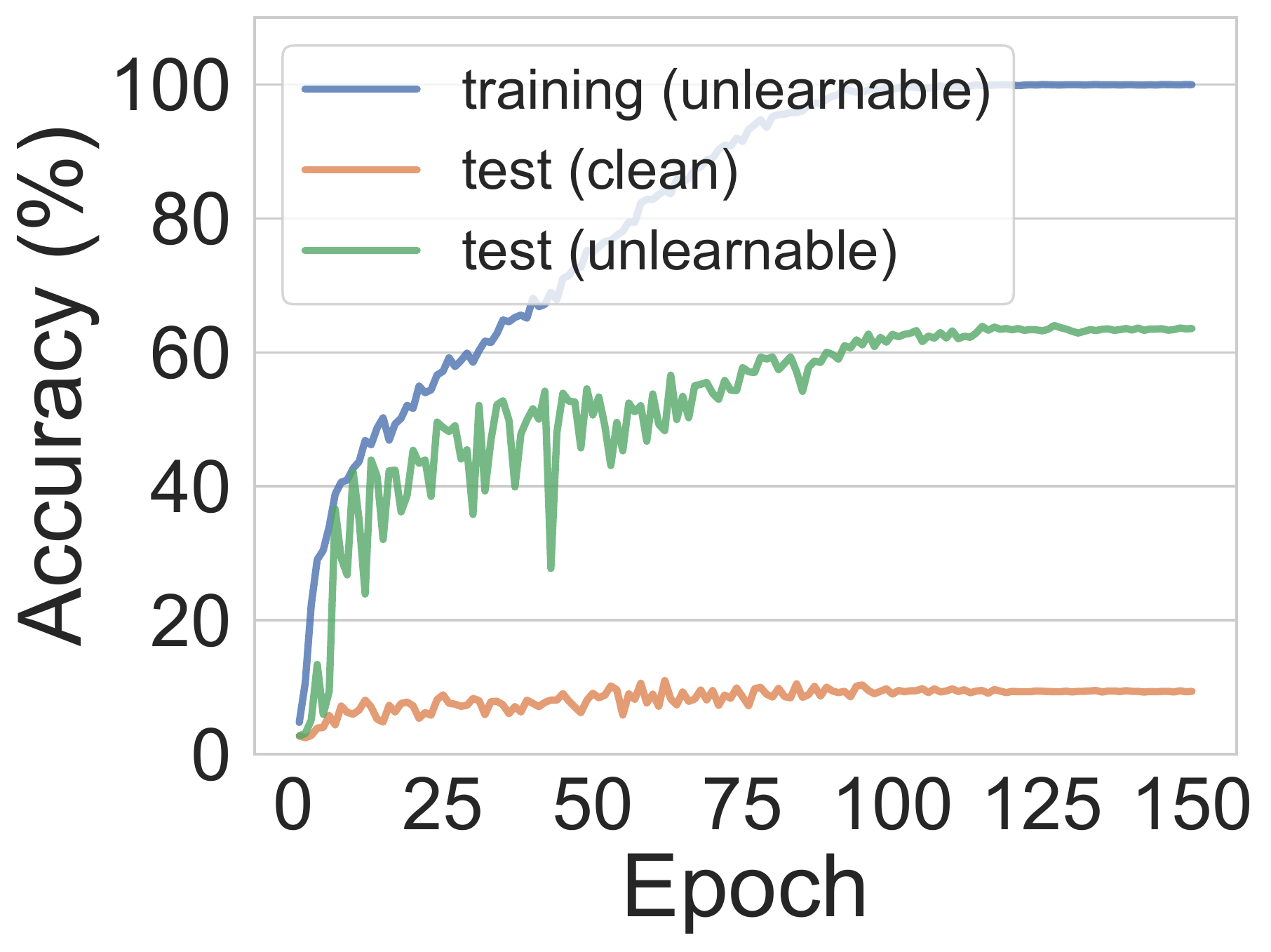}
\centerline{(b) Accuracy trends}
\end{minipage}
\caption{(a) The test accuracy (\%) of ResNet-18 trained on unlearnable-clean mixed vs. clean-only data; and (b) the accuracy trends on clean vs. unlearnable examples. The unlearnable examples are crafted using our UC method on Pets dataset.}
 \label{fig_6}

\end{figure}


All the above experiments are conducted under the assumption that all samples in the dataset are protected, a commonly adopted assumption in the literature~\citep{huang2020unlearnable, fowl2021adversarial, yu2022availability}.
This setting is reasonable when the protectors have the access to the entire dataset, e.g., an online social media company adopts the technique to protect the contents created by all of its users. A more general case is that only a certain proportion of the users protect their data while others do not. This results in mixed dataset with both clean and unlearnable samples.
Here we test our UC method under this setting and show the change in test accuracy with the number of clean classes in Figure~\ref{fig_6} (a). I.e., for the mixture dataset, the rest of the classes are made unlearnable by UC. It can be inferred that the unlearnable classes almost do not contribute to the model training, a similar conclusion as in previous works~\citep{huang2020unlearnable, fowl2021adversarial, yu2022availability}. This implies that only those who adopt the technique will get protected.


\subsection{More Understanding}\label{sec_exp_7}

Why our UCs are more powerful than standard UEs against label-agnostic exploitation? As we explained in Section~\ref{sec:3.1}, the idea of UCs is inspired by the effectiveness of disrupting the uniformity and discrepancy in preventing the model from learning useful information. However, this also raises another question: what exactly does the target model learn? To answer these two questions, here we analyze the learning curves of the target model on the clean vs. unlearnable examples separately. As shown in Figure~\ref{fig_6} (b), as the training progresses, the training accuracy on the unlearnable training samples steadily improves until it reaches $100\%$. But there is almost no improvement in the clean test accuracy on the clean test samples. This is consistent with the the above experimental results that the target model has not learned the capability to perceive normal samples. Surprisingly, however, the model's accuracy on the perturbed test samples is fairly high ($> 60\%$), considering that the normally trained ResNet-18 only achieves a test accuracy of $62.31\%$ on clean Pets dataset. This implies that the unlearnable noise distribution contained in the UCs has effectively concealed the real data distribution.

\section{Conclusion}

Unlearnable examples (UEs) have shown great potential in preventing hackers from using users' private data to train commercial or malicious models. A number of methods have been proposed to improve UEs' transferability and robustness to different datasets, target models and training paradigms. In this work, we identified one limitation of existing UE methods, i.e., their label-consistency assumption. To overcome this limitation, we proposed a more general setting where the hackers could exploit the protected data with different sets of labels. We termed this more challenging setting as \emph{label-agnostic}, and proposed an Unlearnable Clusters (UCs) technique with conditioned generator models, K-means clustering, and large-scale vision-and-language pre-training model CLIP, to craft effective UEs against a wide range of datasets and target models. We also demonstrate its effectiveness against commercial platforms Microsoft {\tt Azure} and Baidu {\tt PaddlePaddle}.

\section{Acknowledgements}

This work is supported by the National Key R\&D Program of China (No. 2021ZD0112804), and Beijing Natural Science Foundation (No.JQ20023), and the Science and Technology Commission of Shanghai Municipality (No. 22511106102), and the National Natural Science Foundation of China (No. U1934220, 61832002, 62172094, 62276067), and the Tianjin Natural Science Foundation (No. 20JCZDJC00400).

{\small
\bibliographystyle{ieee_fullname}
\bibliography{egbib}

\begin{thebibliography}{10}\itemsep=-1pt

\bibitem{biggio2012poisoning}
Battista Biggio, Blaine Nelson, and Pavel Laskov.
\newblock Poisoning attacks against support vector machines.
\newblock {\em arXiv preprint arXiv:1206.6389}, 2012.

\bibitem{biggio2018wild}
Battista Biggio and Fabio Roli.
\newblock Wild patterns: Ten years after the rise of adversarial machine
  learning.
\newblock {\em Pattern Recognition}, 84:317--331, 2018.

\bibitem{bossard2014food}
Lukas Bossard, Matthieu Guillaumin, and Luc~Van Gool.
\newblock Food-101--mining discriminative components with random forests.
\newblock In {\em European conference on computer vision}, pages 446--461.
  Springer, 2014.

\bibitem{chen2020simple}
Ting Chen, Simon Kornblith, Mohammad Norouzi, and Geoffrey Hinton.
\newblock A simple framework for contrastive learning of visual
  representations.
\newblock In {\em International conference on machine learning}, pages
  1597--1607. PMLR, 2020.

\bibitem{chen2017targeted}
Xinyun Chen, Chang Liu, Bo Li, Kimberly Lu, and Dawn Song.
\newblock Targeted backdoor attacks on deep learning systems using data
  poisoning.
\newblock {\em arXiv preprint arXiv:1712.05526}, 2017.

\bibitem{cherepanova2021lowkey}
Valeriia Cherepanova, Micah Goldblum, Harrison Foley, Shiyuan Duan, John
  Dickerson, Gavin Taylor, and Tom Goldstein.
\newblock Lowkey: Leveraging adversarial attacks to protect social media users
  from facial recognition.
\newblock {\em arXiv preprint arXiv:2101.07922}, 2021.

\bibitem{cubuk2018autoaugment}
Ekin~D Cubuk, Barret Zoph, Dandelion Mane, Vijay Vasudevan, and Quoc~V Le.
\newblock Autoaugment: Learning augmentation policies from data.
\newblock {\em arXiv preprint arXiv:1805.09501}, 2018.

\bibitem{devlin2018bert}
Jacob Devlin, Ming-Wei Chang, Kenton Lee, and Kristina Toutanova.
\newblock Bert: Pre-training of deep bidirectional transformers for language
  understanding.
\newblock {\em arXiv preprint arXiv:1810.04805}, 2018.

\bibitem{feng2019learning}
Ji Feng, Qi-Zhi Cai, and Zhi-Hua Zhou.
\newblock Learning to confuse: generating training time adversarial data with
  auto-encoder.
\newblock {\em Advances in Neural Information Processing Systems}, 32, 2019.

\bibitem{fowl2021adversarial}
Liam Fowl, Micah Goldblum, Ping-yeh Chiang, Jonas Geiping, Wojciech Czaja, and
  Tom Goldstein.
\newblock Adversarial examples make strong poisons.
\newblock {\em Advances in Neural Information Processing Systems},
  34:30339--30351, 2021.

\bibitem{fu2022robust}
Shaopeng Fu, Fengxiang He, Yang Liu, Li Shen, and Dacheng Tao.
\newblock Robust unlearnable examples: Protecting data against adversarial
  learning.
\newblock In {\em International Conference on Learning Representations}, 2022.

\bibitem{goodfellow2014explaining}
Ian~J Goodfellow, Jonathon Shlens, and Christian Szegedy.
\newblock Explaining and harnessing adversarial examples.
\newblock {\em arXiv preprint arXiv:1412.6572}, 2014.

\bibitem{gu2017badnets}
Tianyu Gu, Brendan Dolan-Gavitt, and Siddharth Garg.
\newblock Badnets: Identifying vulnerabilities in the machine learning model
  supply chain.
\newblock {\em arXiv preprint arXiv:1708.06733}, 2017.

\bibitem{he2022indiscriminate}
Hao He, Kaiwen Zha, and Dina Katabi.
\newblock Indiscriminate poisoning attacks on unsupervised contrastive
  learning.
\newblock {\em arXiv preprint arXiv:2202.11202}, 2022.

\bibitem{he2016deep}
Kaiming He, Xiangyu Zhang, Shaoqing Ren, and Jian Sun.
\newblock Deep residual learning for image recognition.
\newblock In {\em Proceedings of the IEEE conference on computer vision and
  pattern recognition}, pages 770--778, 2016.

\bibitem{hill2020secretive}
Kashmir Hill.
\newblock The secretive company that might end privacy as we know it.
\newblock In {\em Ethics of Data and Analytics}, pages 170--177. Auerbach
  Publications, 2020.

\bibitem{huang2020unlearnable}
Hanxun Huang, Xingjun Ma, Sarah~Monazam Erfani, James Bailey, and Yisen Wang.
\newblock Unlearnable examples: Making personal data unexploitable.
\newblock In {\em International Conference on Learning Representations}, 2021.

\bibitem{huang2020metapoison}
W~Ronny Huang, Jonas Geiping, Liam Fowl, Gavin Taylor, and Tom Goldstein.
\newblock Metapoison: Practical general-purpose clean-label data poisoning.
\newblock {\em Advances in Neural Information Processing Systems},
  33:12080--12091, 2020.

\bibitem{koh2017understanding}
Pang~Wei Koh and Percy Liang.
\newblock Understanding black-box predictions via influence functions.
\newblock In {\em International conference on machine learning}, pages
  1885--1894. PMLR, 2017.

\bibitem{krause20133d}
Jonathan Krause, Michael Stark, Jia Deng, and Li Fei-Fei.
\newblock 3d object representations for fine-grained categorization.
\newblock In {\em Proceedings of the IEEE international conference on computer
  vision workshops}, pages 554--561, 2013.

\bibitem{krizhevsky2009learning}
Alex Krizhevsky, Geoffrey Hinton, et~al.
\newblock Learning multiple layers of features from tiny images.
\newblock 2009.

\bibitem{li2021align}
Junnan Li, Ramprasaath Selvaraju, Akhilesh Gotmare, Shafiq Joty, Caiming Xiong,
  and Steven Chu~Hong Hoi.
\newblock Align before fuse: Vision and language representation learning with
  momentum distillation.
\newblock {\em Advances in neural information processing systems},
  34:9694--9705, 2021.

\bibitem{li2019visualbert}
Liunian~Harold Li, Mark Yatskar, Da Yin, Cho-Jui Hsieh, and Kai-Wei Chang.
\newblock Visualbert: A simple and performant baseline for vision and language.
\newblock {\em arXiv preprint arXiv:1908.03557}, 2019.

\bibitem{liu2020reflection}
Yunfei Liu, Xingjun Ma, James Bailey, and Feng Lu.
\newblock Reflection backdoor: A natural backdoor attack on deep neural
  networks.
\newblock In {\em European Conference on Computer Vision}, pages 182--199.
  Springer, 2020.

\bibitem{liu2021going}
Zhuoran Liu, Zhengyu Zhao, Alex Kolmus, Tijn Berns, Twan van Laarhoven, Tom
  Heskes, and Martha Larson.
\newblock Going grayscale: The road to understanding and improving unlearnable
  examples.
\newblock {\em arXiv preprint arXiv:2111.13244}, 2021.

\bibitem{madry2017towards}
Aleksander Madry, Aleksandar Makelov, Ludwig Schmidt, Dimitris Tsipras, and
  Adrian Vladu.
\newblock Towards deep learning models resistant to adversarial attacks.
\newblock {\em arXiv preprint arXiv:1706.06083}, 2017.

\bibitem{nilsback2008automated}
Maria-Elena Nilsback and Andrew Zisserman.
\newblock Automated flower classification over a large number of classes.
\newblock In {\em 2008 Sixth Indian Conference on Computer Vision, Graphics \&
  Image Processing}, pages 722--729. IEEE, 2008.

\bibitem{parkhi2012cats}
Omkar~M Parkhi, Andrea Vedaldi, Andrew Zisserman, and CV Jawahar.
\newblock Cats and dogs.
\newblock In {\em 2012 IEEE conference on computer vision and pattern
  recognition}, pages 3498--3505. IEEE, 2012.

\bibitem{poursaeed2018generative}
Omid Poursaeed, Isay Katsman, Bicheng Gao, and Serge Belongie.
\newblock Generative adversarial perturbations.
\newblock In {\em Proceedings of the IEEE Conference on Computer Vision and
  Pattern Recognition}, pages 4422--4431, 2018.

\bibitem{radford2021learning}
Alec Radford, Jong~Wook Kim, Chris Hallacy, Aditya Ramesh, Gabriel Goh,
  Sandhini Agarwal, Girish Sastry, Amanda Askell, Pamela Mishkin, Jack Clark,
  et~al.
\newblock Learning transferable visual models from natural language
  supervision.
\newblock In {\em International Conference on Machine Learning}, pages
  8748--8763. PMLR, 2021.

\bibitem{radosavovic2020designing}
Ilija Radosavovic, Raj~Prateek Kosaraju, Ross Girshick, Kaiming He, and Piotr
  Doll{\'a}r.
\newblock Designing network design spaces.
\newblock In {\em Proceedings of the IEEE/CVF conference on computer vision and
  pattern recognition}, pages 10428--10436, 2020.

\bibitem{ren2022transferable}
Jie Ren, Han Xu, Yuxuan Wan, Xingjun Ma, Lichao Sun, and Jiliang Tang.
\newblock Transferable unlearnable examples.
\newblock {\em arXiv preprint arXiv:2210.10114}, 2022.

\bibitem{russakovsky2015imagenet}
Olga Russakovsky, Jia Deng, Hao Su, Jonathan Krause, Sanjeev Satheesh, Sean Ma,
  Zhiheng Huang, Andrej Karpathy, Aditya Khosla, Michael Bernstein, et~al.
\newblock Imagenet large scale visual recognition challenge.
\newblock {\em International journal of computer vision}, 115(3):211--252,
  2015.

\bibitem{schwarzschild2021just}
Avi Schwarzschild, Micah Goldblum, Arjun Gupta, John~P Dickerson, and Tom
  Goldstein.
\newblock Just how toxic is data poisoning? a unified benchmark for backdoor
  and data poisoning attacks.
\newblock In {\em International Conference on Machine Learning}, pages
  9389--9398. PMLR, 2021.

\bibitem{selim1984k}
Shokri~Z Selim and Mohamed~A Ismail.
\newblock K-means-type algorithms: A generalized convergence theorem and
  characterization of local optimality.
\newblock {\em IEEE Transactions on pattern analysis and machine intelligence},
  (1):81--87, 1984.

\bibitem{shafahi2018poison}
Ali Shafahi, W~Ronny Huang, Mahyar Najibi, Octavian Suciu, Christoph Studer,
  Tudor Dumitras, and Tom Goldstein.
\newblock Poison frogs! targeted clean-label poisoning attacks on neural
  networks.
\newblock {\em Advances in neural information processing systems}, 31, 2018.

\bibitem{szegedy2013intriguing}
Christian Szegedy, Wojciech Zaremba, Ilya Sutskever, Joan Bruna, Dumitru Erhan,
  Ian Goodfellow, and Rob Fergus.
\newblock Intriguing properties of neural networks.
\newblock {\em arXiv preprint arXiv:1312.6199}, 2013.

\bibitem{tan2019efficientnet}
Mingxing Tan and Quoc Le.
\newblock Efficientnet: Rethinking model scaling for convolutional neural
  networks.
\newblock In {\em International conference on machine learning}, pages
  6105--6114. PMLR, 2019.

\bibitem{wold1987principal}
Svante Wold, Kim Esbensen, and Paul Geladi.
\newblock Principal component analysis.
\newblock {\em Chemometrics and intelligent laboratory systems}, 2(1-3):37--52,
  1987.

\bibitem{xiao2010sun}
Jianxiong Xiao, James Hays, Krista~A Ehinger, Aude Oliva, and Antonio Torralba.
\newblock Sun database: Large-scale scene recognition from abbey to zoo.
\newblock In {\em 2010 IEEE computer society conference on computer vision and
  pattern recognition}, pages 3485--3492. IEEE, 2010.

\bibitem{yu2022availability}
Da Yu, Huishuai Zhang, Wei Chen, Jian Yin, and Tie-Yan Liu.
\newblock Availability attacks create shortcuts.
\newblock In {\em Proceedings of the 28th ACM SIGKDD Conference on Knowledge
  Discovery and Data Mining}, pages 2367--2376, 2022.

\bibitem{yun2019cutmix}
Sangdoo Yun, Dongyoon Han, Seong~Joon Oh, Sanghyuk Chun, Junsuk Choe, and
  Youngjoon Yoo.
\newblock Cutmix: Regularization strategy to train strong classifiers with
  localizable features.
\newblock In {\em Proceedings of the IEEE/CVF international conference on
  computer vision}, pages 6023--6032, 2019.

\bibitem{zhang2017mixup}
Hongyi Zhang, Moustapha Cisse, Yann~N Dauphin, and David Lopez-Paz.
\newblock mixup: Beyond empirical risk minimization.
\newblock {\em arXiv preprint arXiv:1710.09412}, 2017.

\bibitem{zhang2020adversarial}
Jiaming Zhang, Jitao Sang, Xian Zhao, Xiaowen Huang, Yanfeng Sun, and Yongli
  Hu.
\newblock Adversarial privacy-preserving filter.
\newblock In {\em Proceedings of the 28th ACM International Conference on
  Multimedia}, pages 1423--1431, 2020.

\bibitem{zhu2019transferable}
Chen Zhu, W~Ronny Huang, Hengduo Li, Gavin Taylor, Christoph Studer, and Tom
  Goldstein.
\newblock Transferable clean-label poisoning attacks on deep neural nets.
\newblock In {\em International Conference on Machine Learning}, pages
  7614--7623. PMLR, 2019.

\end{thebibliography}
}

\end{document}